# Electrically Driven Varifocal Silicon Metalens

Adeel Afridi[1], Josep Canet-Ferrer[1], Laurent Philippet[1], Johann Osmond[1], Pascal Berto[2], Romain Quidant[1,3, *]

[1] ICFO- Institut de Ciències Fotòniques, The Barcelona Institute of Science and Technology, 08860 Castelldefels, Barcelona, Spain

[2] Université Sorbonne Paris Cité, Université Paris Descartes, Neurophotonics Laboratory, CNRS UMR 8250, 45 Rue des Saints Pères, F-75006 Paris, France

[3] ICREA- Institució Catalana de Recerca i Estudis Avançats, 08010 Barcelona, Spain

* romain.quidant@icfo.eu

**Abstract:**
**Optical metasurfaces have shown to be a powerful approach to planar optical elements, enabling an unprecedented control over light phase and amplitude. At that stage, where wide variety of static functionalities have been accomplished, most efforts are being directed towards achieving reconfigurable optical elements. Here, we present our approach to an electrically controlled varifocal metalens operating in the visible frequency range. It relies on dynamically controlling the refractive index environment of a silicon metalens by means of an electric resistor embedded into a thermo-optical polymer. We demonstrate precise and continuous tuneability of the focal length and achieve focal length variation larger than the Rayleigh length for voltage as small as 12 volts. The system time-response is of the order of 100 ms, with the potential to be reduced with further integration. Finally, the imaging capability of our varifocal metalens is successfully validated in an optical microscopy setting. Compared to conventional bulky reconfigurable lenses, the presented technology is a lightweight and compact solution, offering new opportunities for miniaturized smart imaging devices.**

Metasurfaces are 2D metamaterials that have gained enormous attention due to their ability to manipulate light wavefronts at the subwavelength scale, thus enabling planarization of bulky elements.[1–4] Optical metamaterials, in general, are formed by optical scatterers as building blocks, also known as meta-atoms, assembled judiciously to provide operation on demand. Not only these meta-atoms offer planarization, as in case of metasurfaces; they also allow exotic properties, such as cloaking,[5–9] negative refractive index[10–14] and super/hyperlensing,[15–20] which cannot be achieved with conventional optical materials. In recent years, a considerable amount of efforts has been devoted to the design of metalenses.[2,21–28] Based on the geometrical and physical properties, meta-atoms impart a phase, ranging from 0 to 2π, on the impinging light and are arranged according to hyperboloid lens phase profile.[27] Metalenses can be made either plasmonic[21–23,28] or dielectrics.[2,24–27,29] Plasmonics-based metalenses tend to suffer from a low efficiency and transmission owing to the intrinsic losses of metallic scatterers in the visible. Conversely, metalenses based on high susceptibility dielectric materials offer high transmission, efficiency, and compatibility with current microelectronics technology.[30] In addition, both plasmonic and dielectric meta-atoms are also capable of controlling, locally, other properties of light such as polarization, direction and amplitude.

While most progresses so far have focused on static metalens, a further penetration into the industrial sector requires developing reconfigurable metalenses whose focus can be adjusted in real time. In the broad context of metasurfaces, reconfigurability has been achieved by embedding metasurface in liquid crystal,[31,32] electrically driven carrier accumulation/depletion,[33–35] and phase change and phase transition materials including GST,[36–38] $VO_2$,[39–42] $V_2O_3$,[43,44] $SmNiO_3$,[45] and $NdNiO_3$.[46] All these techniques can be potentially employed to design varifocal metalenses. More recent examples of varifocal metalenses are based on a mechanical actuation. For instance, several works have proposed configuration where the metalens stands on a stretchable or elastic substrate allowing the focal length to be changed upon mechanical strain.[47,48] While enabling diffraction limited focusing with large focal distance tuning (> 130 %[47] and > 66 %[48]) and maintaining high efficiency, this approach entails an external mechanical control which may alter the integration and limits the metalens response time. Alternatively, Arbabi and colleagues have recently demonstrated a compact MEMS-based tunable metalens[49] with large optical power tuning and scanning speed reaching up to few kHz. Another strategy recently reported by She *et al.*, relies on electrically tunable dielectric elastomer actuators (DEAs)[50] capable of focal distance tuning of more than 100 %, astigmatism and image shift correction, all simultaneously. While very promising, both configurations require high power/voltage (80 V and 3 kV, respectively) to achieve a substantial focal change.

Here, we present a varifocal metalens formed by cascading a static silicon-on-quartz metalens with electro-thermo-optical control. Focus tuneability is achieved by properly controlling the temperature profile within a thermo-optical material.[51] We fully characterize the system performance, including beam profiling, focus tuneability range and time response. Finally, we demonstrate it suitability for adjustable imaging.

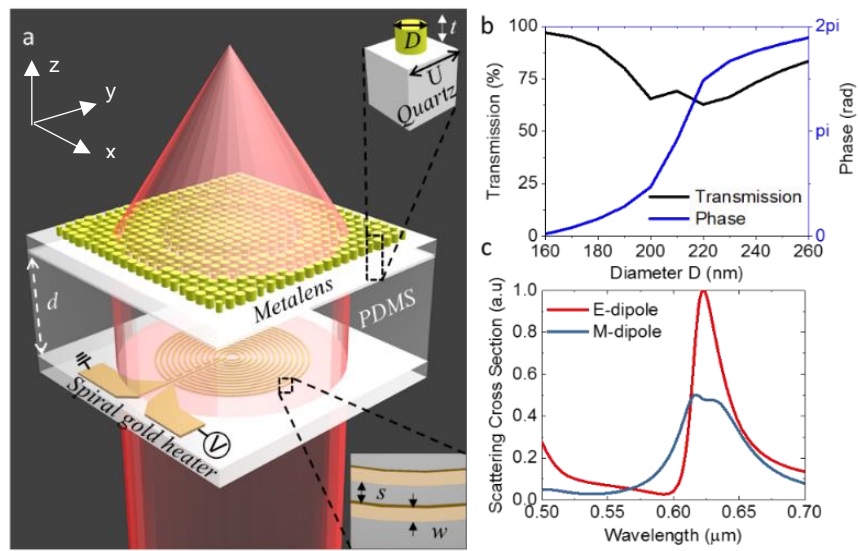

*Figure 1: Concept of electrically-controlled varifocal metalens. (a) Schematic of the system illustrating both building blocks. (b) Simulated optical transmission and accumulated phase at 632 nm for a silicon nano-disk of varying diameter, U = D + 90 nm, and t = 100 nm. (c) Simulated scattering cross-section contribution of electric and magnetic dipole for a silicon nano-disk with D = 210 nm, t = 100 nm, and U = 300 nm.*

Our varifocal system is formed of two building blocks; a high permittivity metalens and an electro-thermo-optical module (Figure 1(a)). The latter, located in the back front of the metalens, enables to dynamically control the optical phase accumulated by the transmitted light before it is focused. The metalens is 300 µm diameter and made of amorphous silicon-on-quartz cylindrical pillars of 100 nm height. The periodicity of the unit cell *U* is equal to *D* + 90 nm, where *D* is the diameter of the silicon nano-disk, optimized to cover the phase from 0 to 2π and satisfy 1st Kerker condition at visible wavelength of 632 nm.[24,52] Optimal diameters for the silicon nano-disks were identified by using the commercial finite element method (FEM)-based simulation tool COMSOL MULTIPHYSICS (See Supporting Information for detail). Figure 1(b) shows the simulated transmission and phase of silicon nano-disks for diameters *D* ranging from 160 nm to 260 nm, at 632 nm wavelength. In addition, we also identified that for a disk diameter of 210 nm, electric and magnetic dipole resonances overlap, enabling to minimize reflection (Figure 1(c)). The scattering cross-section contribution given in Figure 1(c) were calculated using multipole decomposition.[24] To form the metalens, silicon nano-disks were arranged according to the following hyperboloid phase profile[27]:

$$\phi_t(x,y) = 2\pi - \frac{2\pi}{\lambda}\left(\sqrt{x^2 + y^2 + f^2} - f\right) \qquad 1$$

Where $\phi_t(x,y)$ stands as the required phase profile for the nano-disk placed at coordinates (*x, y*), *f* is the focal length, and λ is the wavelength of operation (632 nm).

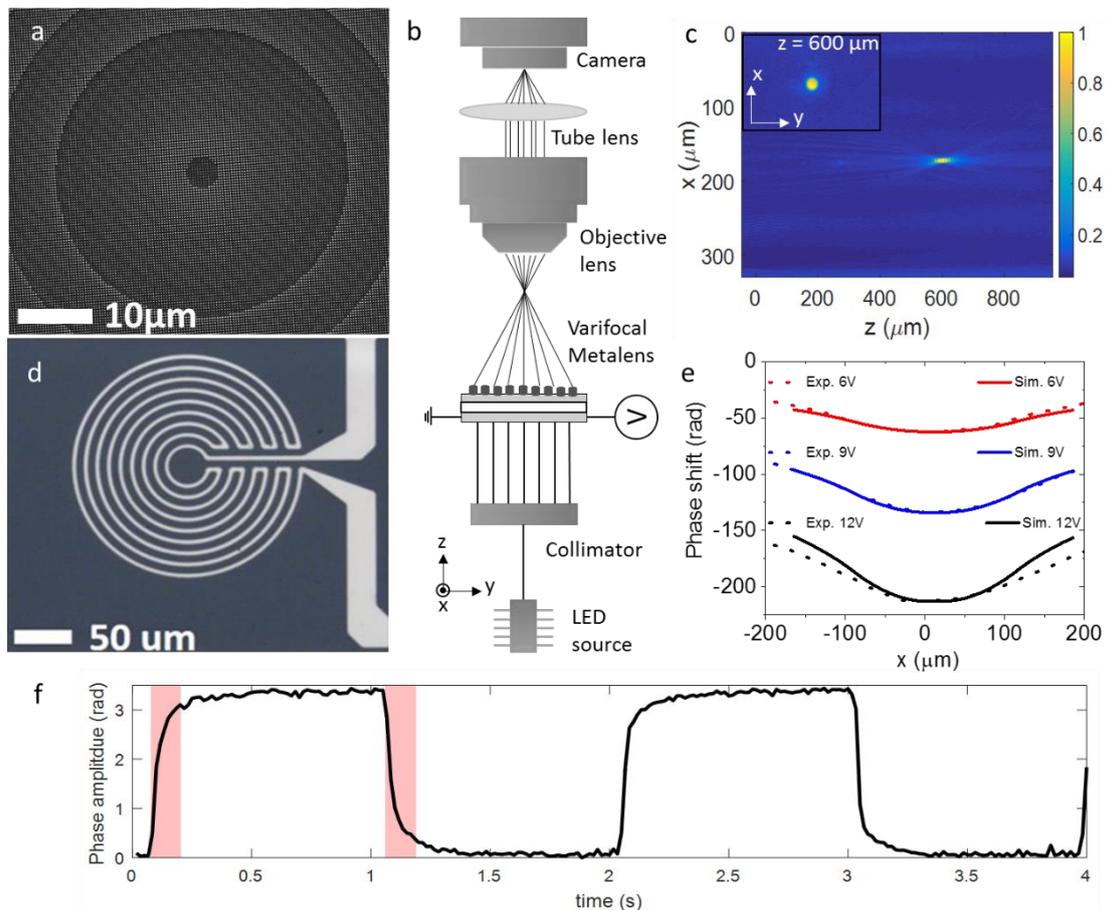

Figure 2: Description and characterization of the fabricated varifocal metalens. (a) SEM image of the central portion of the silicon metalens. (b) Optical setup for the characterization of the varifocal metalens. (c) Optical mapping of the focus created by a silicon metalens with diameter of 300 µm and focal length of 600 µm (without gold spiral heater). The inset shows the spot at focal plane (z = 600 µm) and the color bar refers to normalized intensity. (d) SEM image of the gold spiral heater. (e) Simulated (solid lines) and experimental (dashed lines) thermally induced phase shift (in radians) by gold spiral heater with diameter

*of 200 μm, as a function of applied voltage. (f) Time response of gold spiral heater for applied square wave signal (0 - 2.5 V). The rise and fall time are highlighted in red areas.*

The electro-thermo-optical module consists of a gold spiral resistor of 200 μm diameter covered with a PDMS slab of thickness *d* equal to 700 μm (Figure 1(a)). The spiral geometry was chosen in order to introduce a paraboloid shaped refractive index gradient in the PDMS layer. The gold stripes are 4 μm wide, 50 nm thick and spaced by 6 μm, as shown in Figure 1(a) (bottom right inset). We fabricated the electrically varifocal metalens using combination of electron beam lithography and UV lithography technique (see Supporting Information). Figure 2(a) shows the scanning electron microscopy (SEM) image of the central portion of the fabricated metalens with focal length of 600 μm and Figure 2(d) shows an electron micrograph of the fabricated gold spiral heater.

For the characterization of the varifocal metalens, we used a homemade optical setup, sketched in Figure 2(b). A 632 nm LED light source is collimated on the varifocal metalens which is placed on a 3D micrometric stage. All the light transmitted through the varifocal metalens is focused onto an intermediate plane, and re-imaged onto a CCD camera by a custom-built microscope, consisting of a microscope objective (x20 magnification) and a tube lens. DC power supply is used to apply voltage to the spiral resistor. For reference, we first characterized the metalens (theoretically designed to have a focal length of 600 μm) in absence of gold spiral resistor. Mapping of the optical intensity (Figure 2(c)) shows the experimental focal length to be exactly 600 μm and the focal spot diameter of 7.5 μm.

Subsequently, we also simulated and experimentally characterized the PDMS coated gold spiral heater (without metalens) in order to quantify its effect on the incoming wavefront (see Supporting Information). Simulations were performed for the same applied voltages of 6, 9 and 12 V used in our experiments. We extracted the temperature profile induced into the PDMS from the simulation and calculated the associated refractive index using the following equation:

$$n(T) = n(T_0) + \left(\frac{dn}{dT}\right) * (T - T_0) \qquad\qquad 2$$

Where, $n(T)$ is the temperature dependent refractive index of PDMS, $n(T_0)$ is the refractive index of PDMS at room temperature and is equal to 1.412,[53] $\left(\frac{dn}{dT}\right)$ is equal to -4.5*10$^{-4}$ (1/K)[54] and $T_0$ is the room temperature in Kelvin, i.e. 298 K. Finally, we calculate the thermally induced phase-shift, i.e. the difference in optical path of an incoming planar wave associated to switching on the heater, by integrating the refractive index along the axial direction (*z*-axis).

The numerical simulations are compared to experimental wavefront measurements performed on the electro-thermo-optical module (Figure 2(e)). The simulated (solid lines) and experimental phase shift profiles along the *x*-axis (dashed lines) show an excellent agreement for applied voltages of 6 and 9 V while further deviation is observed for 12 V.

Once validated the phase profile introduced by the electro-thermo-optical module, we are interested in its time response, which will entirely dictate the operation speed of the varifocal lens. Figure 2(f) shows the maximum amplitude of the thermal phase-shift plotted against time, upon a periodic electrical driving with a square signal at 0.5 Hz and 2.5 V amplitude. Our data give a rise time of 125 ms (10 % to 90 % rise) and a fall time of 130 ms (90 % to 10 % fall) highlighted by the red shedding. These values are in good agreement with the time scale of the temperature evolution[55] $\tau = l_c^2/(4Ds) \approx 100\ ms$, where $l_c$ is the characteristic length of the system, i.e. the spiral diameter ($l_c$ = 200 μm), and $Ds$ the thermal diffusivity of the surrounding ($Ds \approx 10^{-7}\ m^2.s^{-1}$ for PDMS). It is noteworthy mentioning that this characteristic time could be significantly reduced by decreasing the spiral diameter $l_c$. For instance, for a spiral diameter $l_c$ = 20 µm, we expect the response time to decrease to $\tau \approx 1\ ms$.

In addition, we also measured the transmittance of the metalenses and the gold spiral heater, separately. For this purpose we used the same optical setup as in Figure 2(b), replacing the camera by a fiber-coupled spectrometer (SHAMROCK - SR-303I-A). By using a pinhole, we made sure that the light coupled to the spectrometer only originated from an aperture size equal to the diameter of metalens or gold spiral heater. The transmittance at 632 nm of metalenses with 600 µm and 1000 µm focal length are listed in Tables 1 and 2, respectively. For the spiral gold heater module the transmittance was measured as a function of the applied voltage, being 73 %, 72 %, 71 % and 69 % for applied voltages of 0, 3, 6, and 9 V, respectively. A transmittance decrease with increasing voltage is expected due to the increase in temperature which leads to increasing absorption losses in gold.

Other important parameters to characterize our device, are the focusing efficiency and Strehl ratio. We determined the focusing efficiency by dividing the amount of light passing through an aperture of radius equal to *3 times the FWHM* of the focal spot by the total input light illuminating the metalens.[26] To calculate the Strehl ratio, we followed the same protocol as given by Khorasaninejad *et. al*[56] (also see Supporting Information for details). Both focusing efficiency and Strehl ratio are shown in Table 1 and Table 2 for both of the metalenses.

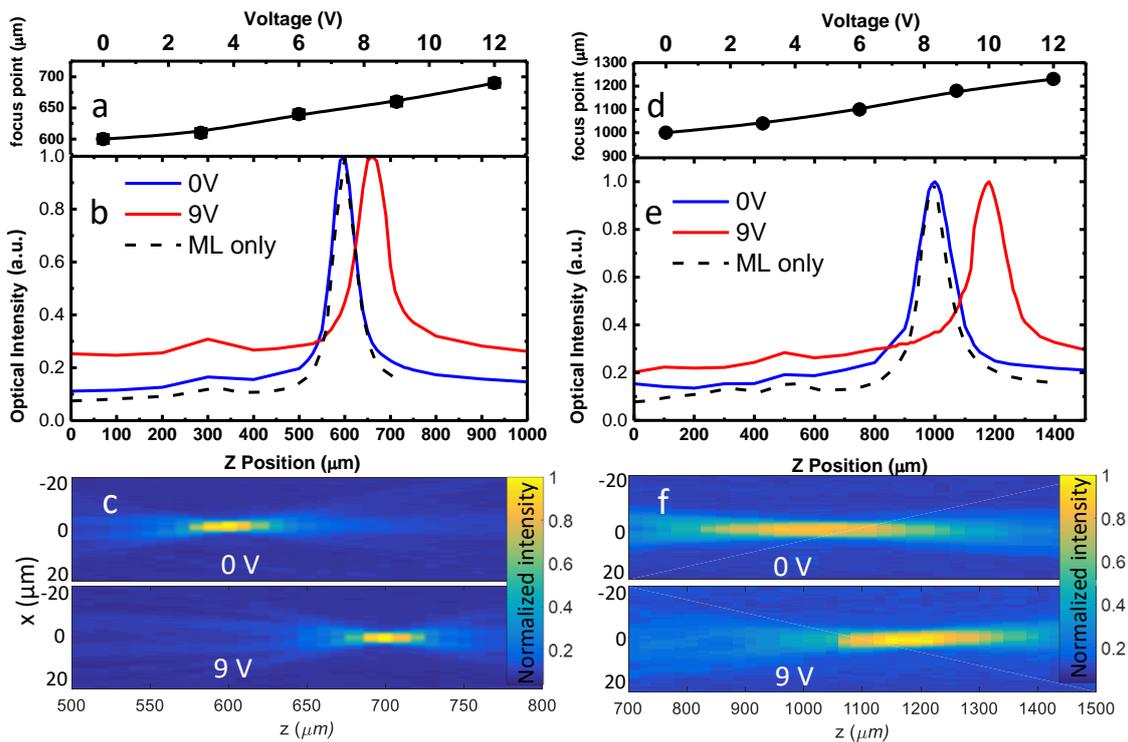

*Figure 3: Tuning capability of the varifocal metalens. (a) – (c) refer to the lens with designed focal length of 600 µm and (d) – (f) show the lens with 1000 µm focal length. (a) and (d) Change in the focal lengths vs different voltages. (b) and (e) Axial (z-crosscut) optical intensity distribution comparison of metalens only and varifocal metalens for 0 and 9 V. (c) and (f) XZ map of optical intensities for 0 and 9 V. The scale bar is with arbitrary units.*

Now the two building blocks have been characterized separately, we are interested in demonstrating focal length tuning by cascading the metalens with the electro-thermo-optical module. Figures 3(a) and 3(d) show the focal length tuneability, as a function of the applied voltage, for lenses with focal lengths of 600 µm and 1000 µm, respectively.

For both lenses, the focal length gradually increases as the applied voltage rises (Figures 3(a) and (d)), following a quasi-linear dependence. At 12 V, the focal length has changed by 15 % and 23 %, for the designed 600 µm and 1000 µm lenses, respectively. This corresponds to a focus change that is larger than the actual Rayleigh length of the metalenses (70 µm and 225 µm, respectively). Beyond the efficient control over the focus under low/moderate voltages, we are interested in assessing any potential aberration the electro-thermo-optical module could introduce in the metalens point spread function. To this aim, we map the XZ intensity distribution for 0 and 9 V (Figures 3(c) and (f)) and extract from them the axial focus profiles (Figures 3(b) and (e)). For reference, we also plot the focal profile for the metalens only (without the heating module). Remarkably, the focus profile is not dramatically altered by the thermo-optical control, even between the two extremities of the explored driving voltage range. Still, one can notice some slight changes in the focus size that are not foreseen to be critical for fine refocusing, for instance in an imaging setting.

We measured the overall transmittance, focusing efficiency and Strehl ratio of the whole device (metalens and gold spiral combined) throughout the all tuning range. Tables 1 and 2 summarize the results for the varifocal metalenses with 600 and 1000 µm, respectively. For both metalenses, the Strehl ratio remains similar as compared to the case of metalens only, indicating the gold spiral heater does not introduce significant aberrations. On the other hand, the overall transmittance slightly decreases with increasing voltage. Similarly, the focusing efficiency slightly diminishes with the inclusion of the gold spiral heater and with increasing voltage. It is worth mentioning that the effect of the thermo-optical module on the transmittance and focusing efficiency can be further minimized by using ITO instead of gold (see Supporting Information).

*Table 1: Characterization of the metalens with 600 µm designed focal length*

| **Metalens only** | | **Transmittance (%)** | **Focusing efficiency (%)** | **Strehl ratio** |
|---|---|---|---|---|
| | | 69 | 76 | 0.72 |
| **Varifocal metalens** | Applied voltage | | | |
| | 0 V | 50.4 | 69.5 | 0.69 |
| | 3 V | 49 | 68.6 | 0.7 |
| | 6 V | 49 | 68 | 0.7 |
| | 9 V | 46 | 66.5 | 0.69 |

*Table 2: Characterization of the metalens with 1000 µm designed focal length*

| **Metalens only** | | **Transmittance (%)** | **Focusing efficiency (%)** | **Strehl ratio** |
|---|---|---|---|---|
| | | 66 | 67 | 0.69 |
| **Varifocal metalens** | Applied voltage | | | |
| | 0 V | 48.1 | 61.1 | 0.67 |
| | 3 V | 47.5 | 59 | 0.67 |
| | 6 V | 46 | 56.7 | 0.65 |
| | 9 V | 45.5 | 55 | 0.66 |

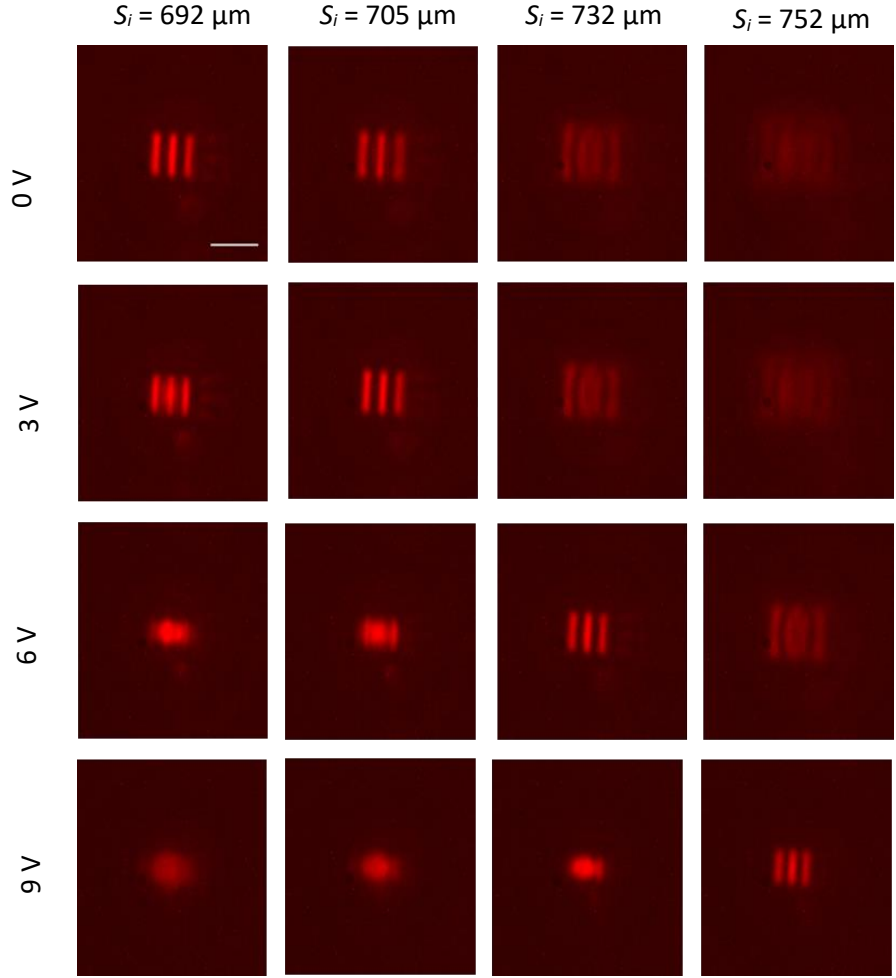

*Figure 4: Imaging experiments performed with the 600 μm varifocal metalens. The scale bar is 10 μm. At 0 V, the image is focused at a plane with distance $S_i$ equal to 692 μm. Additionally, at the same plane we take images for 3, 6, and 9 V to show the image focus shift (1st column). In similar manner, we observe image focus at 705 μm, 732 μm and 752 μm for 3, 6, and 9 V, respectively.*

Finally, we validate the capability of our varifocal metalens for imaging applications, using a commercially available negative 1951 USAF resolution target (R1DS1N, Thorlabs). For this purpose, we use the same optical setup described in Figure 2(b), with little modifications. The resolution target followed by an aspheric lens was placed in between the collimator and varifocal metalens (Figure 2(b)). In this way, the resolution target is first imaged by the aspheric lens and the varifocal metalens in an intermediate image plane, and re-imaged by our custom-built microscope on the camera. In fact our imaging system is similar to the one used by Arbabi *et al.*[49] However, here we kept the distance between object and aspheric lens constant and we move the objective lens farther away from the varifocal metalens to compensate for the shift in the focal length at different voltages. We denote the distance between the varifocal metalens surface and the image by $S_i$. Figure 4 summarizes the imaging data. When no voltage is applied, the image is focused at a distance $S_i$ equal to 692 μm. By switching on the electro-thermo-optical control, the image becomes out of focus. For an applied voltage of 3 V, the image focus moves to a distance of 705 μm. Similarly, results for 6 and 9 V are also given. It is worth noticing that the image planes are well separated from one voltage to another or in other words, that the varifocal metalens shows good resolving capabilities between different planes. Moreover, we performed imaging with the metalens only (without any heater) and compared it with the varifocal metalens (with heater) for 0 V and results are presented in Figure S2 of Supporting information.

In conclusion, we presented an electrically-controlled varifocal metalens based on the thermo-optical effect in PDMS. We achieve focal change higher than the Rayleigh length, with voltages as low as 12 V. Our device driving time response is in the 100 ms range but with the potential to reach to 1 ms by suitable engineering of the electro-thermo-optical module. We note that the electro-thermo-optical module in our device does not introduce significant perturbation to the incoming beam. In its current state, the overall transmission of our device is limited by the transmission of the gold based electro-thermo-optical module. However, by using ITO instead of gold, the overall transmission would be substantially improved. The device shows good performance for imaging making it potentially relevant to adaptive vision, bioimaging, wearable technology and displays.

**Acknowledgments:** This project has received funding from the European Union's Horizon 2020 research and innovation programme under the Marie Skłodowska-Curie grant agreement No 665884. The authors also acknowledge financial support from the European Commission through ERC grant QnanoMECA (CoG - 64790), grant PHENOMENON (780278), Fundació Privada Cellex, CERCA Programme / Generalitat de Catalunya, and the Spanish Ministry of Economy and Competitiveness through the Severo Ochoa Programme for Centres of Excellence in R&D (SEV-2015-0522), grant FIS2016-80293-R, and Juan de la Cierva grant (IJCI-2015-25438). The authors also wish to thank José Garcia Guirardo and Johann Berthelot for their help on the electric resistor fabrication, Philipp del Hougne for preliminary COMSOL simulations as well as Irene Alda, Renaud Marty and Jon Donner for their preliminary work on the electric resistor characterization.

**Supporting Information:**

Simulation of silicon nano-disks and gold spiral heater. Fabrication of the metalens and gold spiral heater. Strehl ratio calculation. Qualitative assessment of imaging of the metalens with and without heating module. Fraunhofer propagation of varifocal metalens with gold and ITO spiral heater.

# Supporting Information
# Electrically Driven Varifocal Silicon Metalens

Adeel Afridi[1], Josep Canet-Ferrer[1], Laurent Philippet[1], Johann Osmond[1], Pascal Berto[2], Romain Quidant[1,3,*]

[1] ICFO- Institut de Ciències Fotòniques, The Barcelona Institute of Science and Technology, 08860 Castelldefels, Barcelona, Spain

[2] Université Sorbonne Paris Cité, Université Paris Descartes, Neurophotonics Laboratory, CNRS UMR 8250, 45 Rue des Saints Pères, F-75006 Paris, France

[3] ICREA- Institució Catalana de Recerca i Estudis Avançats, 08010 Barcelona, Spain

* romain.quidant@icfo.eu

Number of Pages: 6

Number of Figures: 4

### Simulation of silicon nano-disks and gold spiral heater

The RF module of COMSOL MULTIPHYSICS was used to solve the full fields. The simulation domain consists of a single unit cell comprising of semi-infinite quartz substrate, a disk and semi-infinite air medium surrounding the disk. In the x and y direction, periodic boundary conditions were applied while perfectly-matched-layers (PML) were used in z-direction to mimic the semi-infinity of the two mediums. The unit cell was illuminated with a plane wave source propagating in the z-direction. Two ports were defined in the z-direction just before the PML layers, one as source and other as listening port. From these two ports, the S-parameters were derived from which the transmission and phase of the nano-disks were calculated. For accuracy, a very fine mesh of step size less than 5 nm was used for the silicon disks. Refractive index of silicon was obtained from ellipsometric measurements of the silicon samples.

For gold spiral heater simulation, we used Heat Transfer module of COMSOL MULTIPHYSICS. In this case, our domain consists of gold heater sandwiched between glass (substrate) and PDMS. In x and y directions the domain is set ten times bigger than the gold heater, while in z direction (top boundary of PDMS and lower boundary of glass) convective heat flux boundary condition has been used with natural convection of air. We considered the temperature dependent resistivity of the gold. Thermal conductivity, density and heat capacity of PDMS and glass were taken from Polymer data handbook.[1]

### Fabrication of the metalens and gold spiral heater module

For the fabrication of the electrically varifocal metalens, we used commercially available amorphous silicon on quartz substrates (obtained from Siegert Wafer) with a silicon layer thickness of 100 nm. The hyperboloid lens phase profile (equation 1), discretized with periodicity U and appropriate disk diameter given by Figure 1(b), was transferred to a silicon sample by using electron beam lithography (E-beam) technique using negative photoresist (AR-N 7720), followed by reactive ion etching with SF6 and C4F¬8. Finally, the remaining residual resist was cleaned by piranha lift-off. Lenses with two different focal lengths (fixed diameter of 300 µm), 600 µm and 1000 µm were fabricated.

$$\phi_t(x,y) = 2\pi - \frac{2\pi}{\lambda}\left(\sqrt{x^2 + y^2 + f^2} - f\right) \qquad 3$$

The fabrication of the gold spiral heater was a two-step process. We started with fabricating the large electrodes using UV lithography technique on a positive resist (AZ ECI 3027) coated glass. After exposure and development, 50 nm gold was evaporated on the sample (using thermal evaporation) with 2 nm of titanium (using electron beam vapor deposition) as adhesion layer between the gold layer and the glass substrate, before lift-off. As second step, to fabricate the gold spiral, the electrode sample was coated with a positive electron-sensitive resist (SML300). As previously, 2 nm of titanium and 50 nm of gold were deposited over the e-beam exposed and developed sample. Finally, the whole sample with gold spiral heater was coated with 700 µm PDMS and assembled to each of the metalens.

### Strehl ratio calculation

The Strehl ratio was calculated by comparing the measured intensity distribution of focal spot to a calculated profile of Airy-disk for a lens with same NA as our device. The Strehl ratio was defined to be the ratio of peak value of measured intensity to the theoretical Airy-disk. Figure S1 is the example of Strehl ratio calculation for the metalens with designed focal length of 600 µm, with (0 V) and without heating module.

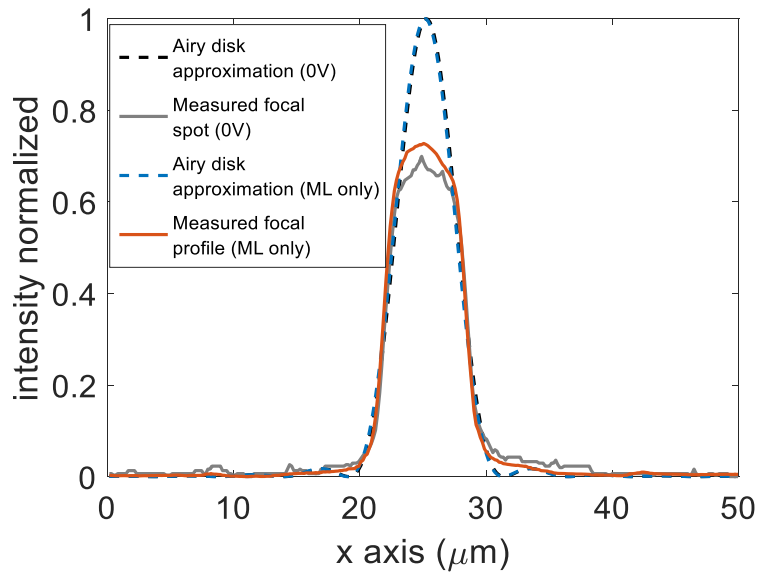

*Figure S5: Example comparison of the airy Airy disk and the measured focal spot. The blue dashed line and red solid line refers to the case of metalens only (without heater) while the black dashed line and grey solid line refer to the varifocal metalens at 0 V. The designed focal length is 600 µm. The peak ratio of the measured focal profile to the Airy disk approximation was taken to calculate the Strehl ratio.*

## Qualitative assessment of imaging of the metalens with and without heating module

For a qualitative assessment we performed imaging experiments with the metalens with 600 µm focal length with and without the heating module. A negative 1951 USAF resolution target (R1DS1N, Thorlabs) was used as a reference object. Figure S2 shows the results of this comparison.

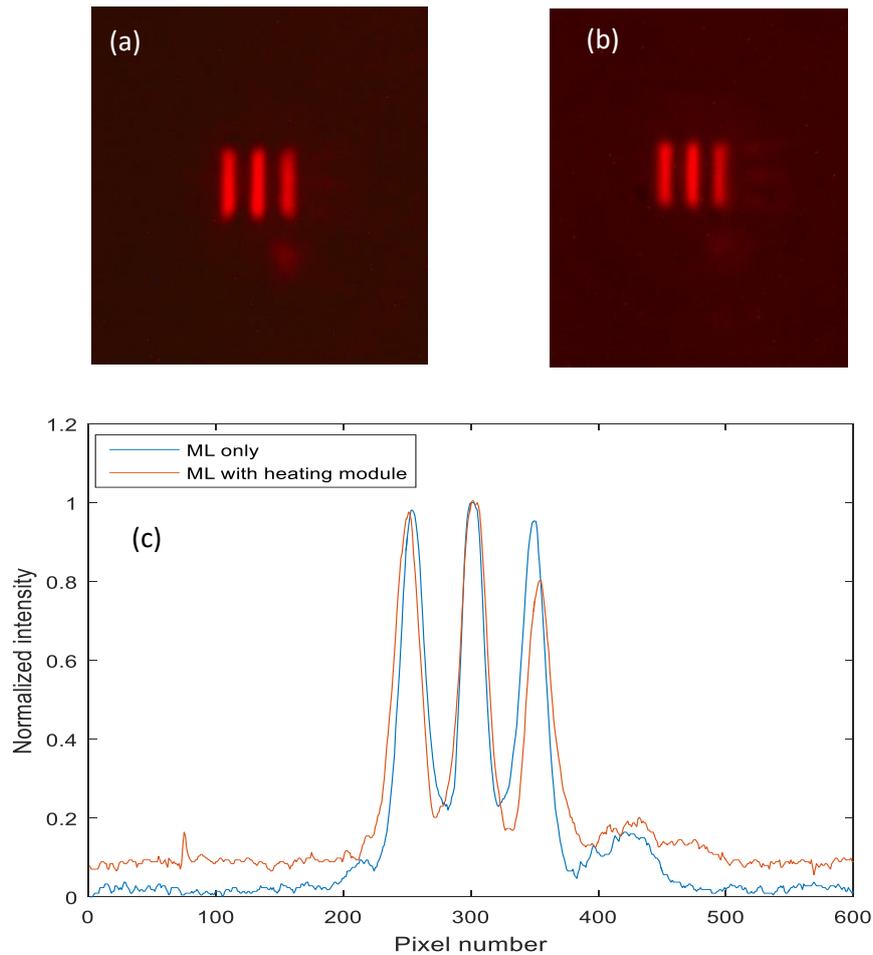

*Figure S2: Qualitative analysis of the imaging capability and aberrations (a) Metalens only with 600 µm focal length. (b) Metalens (600 µm focal length) with gold spiral heater. (c) x- cross cut comparison of both images.*

## Fraunhofer propagation of varifocal metalens with gold and ITO spiral heater

We theoretically simulated the Fraunhofer diffraction[2] of the metalens, with and without the gold spiral heater, in order to compare the PSF in both cases. To this aim, we modelled the metalens by an ideal lens with the same NA, and assumed that the heater is located at the entrance pupil of the metalens and acts as a complex (phase and amplitude) mask. We then modelled the complex transmission of this mask by considering the geometry of the gold spiral heater and the refractive index, absorption and thickness of gold.[3] We imposed this complex mask as an aberration to the ideal metalens. Figure S3 shows the comparison of the calculated PSF with and without heater.

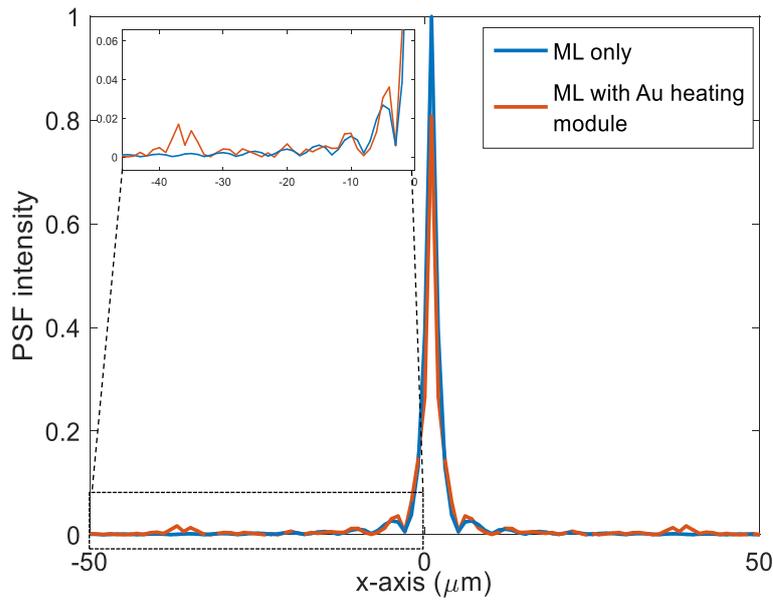

*Figure S3: Comparison of the simulated PSF of a metalens of 600 µm focal length and diameter of 300 µm without (blue solid line) and with gold spiral heater of size 200 µm (red solid line). Inset shows a zoomed region of the black rectangle.*

To complete our study, we also simulated the case where gold was replaced by ITO (same geometrical parameters) as a possible strategy to further decrease the beam perturbation. In this case, aberrations are strongly reduced since ITO induces lower absorption losses (Figure S4).

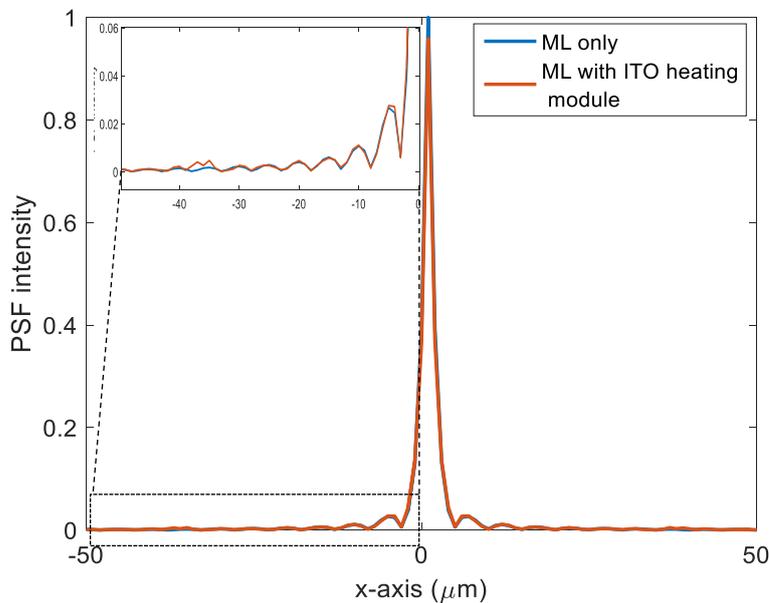

*Figure S4: Comparison of simulated PSF of an ideal metalens only with 600 µm focal length and diameter of 300 µm (blue solid line) and same ideal metalens with ITO spiral heater of size 200 µm. Inset shows the energy that is located outside the central ring of the PSF when the metalens is combined with the gold spiral heater.*